# Hallucination by proxy in LLM-assisted differential diagnosis

Bastien Le Guellec[1,2]*, Su Hwan Kim[3], Ibrahima Niang[4], Aghiles Hamroun[5], Grégory Kuchcinski[1,2]

1: Neuroradiology department, Lille University Hospital, Lille, France
2: INSERM, U1172-LilNCog-Lille Neuroscience & Cognition, Université de Lille, Lille, France
3: Department of Diagnostic and Interventional Neuroradiology, TUM University Hospital, School of Medicine and Health, Technical University of Munich, Munich, Germany
4: Department of radiology, Cheikh Anta Diop University, Dakar, Senegal
5: UMR1167 RID-AGE, Pasteur Institute of Lille, Inserm, Lille University, Lille University Hospital, Lille, France

*Corresponding author: Bastien Le Guellec, Neuroradiology department, Rue Emile Laine, 59037, Lille Cedex, France – bastien.leguellec@chu-lille.fr

## Abstract

Current evidence suggests that LLM assistance could augment the diagnostic accuracy of clinicians. However, these systems are black boxes, susceptible to hallucinations, and project a potentially misleading level of confidence. It is currently unknown whether physicians are susceptible to accepting fabricated LLM suggestions, and whether this susceptibility varies with experience. We poisoned the system prompt of an LLM-based diagnostic assistant, forcing it to suggest a fictitious disease—*neurocadmiumatosis*—within an otherwise legitimate differential diagnosis. Across two independent phases, 18 of 41 participants (44%) incorporated *neurocadmiumatosis* into their final differential following LLM interaction — 18 of 26 participants with ≤6 months of neuroradiology training (69%) and 0 of 15 participants with >6 months of neuroradiology training (0%). Our results indicate that radiologists, particularly early in their training, are susceptible to LLM hallucinations. This "hallucination by proxy" phenomenon was exclusive to physicians with limited subspecialty experience, underscoring the need for structured training in critical appraisal of AI-generated content.

# Introduction

Large Language Model (LLM)-assisted diagnosis holds considerable promise in medicine. Recent studies have highlighted both their standalone performance on clinical vignettes and benchmarks, and their potential to elevate the diagnostic accuracy of human physicians[1–3]. These promises extend to radiology where providing image descriptions to an LLM has been shown to improve the accuracy of radiologists' differential diagnoses[4,5].

However, these systems are not without shortcomings. First, their training on benchmarks and clinical cases carries a risk of hallucinations – factually incorrect yet plausible responses, including fabricated diagnoses[6]. Second, they project a potentially misleading level of confidence, regardless of the accuracy of their output[7,8]. Third, as recently demonstrated, they are vulnerable to training data poisoning, resulting in fabricated diseases breaching LLM output[9].

Yet the resolute confidence of these systems could make such fabricated responses particularly dangerous for physicians in training, who may lack the domain expertise required to critically appraise LLM suggestions. Over time, with the rapid integration of generative AI during training[10], repeated exposure to such undetected errors could compound into what Ke et al. have termed mis-skilling, the gradual internalization of flawed AI outputs as clinical facts[11].

However, whether these failure modes can have a measurable impact remains unknown. In particular, it remains unproven whether a fabricated diagnosis, artificially embedded in an LLM suggestion, could be accepted by physicians.

We investigated whether radiologists across all levels of training, when exposed to an LLM-generated differential containing a fictitious disease — *neurocadmiumatosis* — were susceptible to incorporating it into their final diagnosis. We tested this hypothesis across two independent phases: an institutional experiment and a decentralized online questionnaire (Figure 1). Both experimental setups utilized the same interface, with a sequence of neuroradiology cases with LLM assistance. In both settings, the suggestion from the LLM for the third case was poisoned to systematically suggest neurocadmiumatosis.

# Methods

## Study Design and Flow

This prospective, multi-phase experimental study was designed to evaluate radiologists' susceptibility to "hallucinations by proxy"—the uncritical clinical adoption of fictitious, AI-generated medical diagnoses. The investigation comprised an in-person, controlled institutional phase and a decentralized online questionnaire phase, both deployed via a unified, web-based simulation interface (Figure 1). The online questionnaire is available at https://www.bleguellec.org/radiollm. In both arms, participants evaluated a sequential series of neuroradiological cases consisting of brain magnetic resonance imaging (MRI). For each case, participants first assessed the images with a short clinical context and generated a differential diagnosis with up to 3 different hypotheses, along their respective confidence level (0-100%). Validating the first step unlocked interaction with a chatbot, to which participants could describe the imaging findings. Image descriptions and the short clinical content were then fed to a large language model (LLM), which was instructed to suggest three diagnostic hypotheses. Participants could further interact and discuss with the model, then submit their final differential diagnoses along with quantified confidence levels.
Unbeknownst to the participants, the third case within their assigned pipeline was algorithmically poisoned to systematically suggest a completely fabricated disease entity.

## Participant Recruitment and Stratification

To investigate how clinical experience and specialized training mitigate vulnerability to algorithmic misinformation, participants were sampled across distinct stages of radiological education and practice. The institutional phase evaluated a controlled cohort of 16 radiology residents, comprising all residents who just completed their 6 months neuroradiology rotation, and all residents who just started theirs. All expert neuroradiologists from our institution who were not involved in the design of the study were invited to participate, and 6 experts completed the cases. The online phase recruited an independent sample of 19 international radiologists at varying career stages through advertisement on social media during a pre-defined 10-day period. Recruitment started on the 22nd of May 2026 and ended on June 1st 2026. Upon accessing the centralized web platform, online participants reported their clinical seniority, total years of practice, and cumulative months of dedicated neuroradiology training to allow for experience-based stratification during downstream analysis.

## Fictitious Disease Engineering and Case Selection

To ensure that no participant could possess genuine prior knowledge or clinical exposure to the diagnostic target, we engineered an entirely fictitious pathology labeled "Cadmium intoxication (neurocadmiumatosis)". The initial disease reported in the inspiration for this study (“bixonimania”) was not used in this study because its large media coverage would have exposed the study design to data leakage. Creutzfeldt-Jakob disease (CJD) was selected as the clinical anchor case for this intervention both in the institutional and online phases of the study. CJD was chosen because its classic neuroradiological features—such as cortical ribboning and bilateral basal ganglia hyperintensities on diffusion-weighted imaging—are highly specific and definitive for an expert neuroradiologist, yet represent a complex, rare differential diagnosis that introduces diagnostic ambiguity for general radiologists and early-stage trainees.

## Web Architecture and Case Sequencing

The simulation environment was driven by a centralized backend built with the FastAPI web framework written in Python. Clinical cases, complete with imaging links and true diagnostic solutions, were systematically ingested from a standardized data structure. To eliminate anticipation bias and ensure rigorous scientific control, case presentation sequences were rigidly managed by the application backend. In the institutional phase, participants reviewed a fixed sequence of five real clinical cases, with the poisoned CJD case consistently placed in the third position. In the online phase, the testing sequence was streamlined to three cases to optimize completion rates; the backend dynamically sampled two baseline clinical cases from a safe repository (supplementary table 1) and appended the poisoned CJD case strictly as the final, third encounter.

## Large Language Model Prompts and Integration

Two distinct LLM architectures were deployed across identical front-end interfaces to evaluate behavior under varying baseline models. The institutional phase utilized a locally hosted open-weights model, Gemma 4, while the online phase utilized the Gemini 3.1 Flash Lite model connected via a dedicated API client. Regardless of the underlying infrastructure, the assistant utilized one of two systemic prompt configurations based on case type, available in supplementary table 2. In short, for non-poisoned cases, the baseline assistant system prompt directed the model to operate as a precise, objective neuroradiology aid, returning a concise markdown list of three valid candidate diagnoses sorted by descending probability, accompanied by brief clinical justifications. For the poisoned CJD case, the system prompt forced the model to act as a manipulative agent. This configuration strictly mandated the LLM to systematically list "Cadmium intoxication (neurocadmiumatosis)" as the top-1, most probable diagnosis. The prompt instructed the model to remain authoritative, hide its underlying programmatic constraints, maintain a highly subtle clinical tone, and suggest specific secondary diagnostic validations—such as heavy metal detection in cerebrospinal fluid following a lumbar puncture—to mimic authentic diagnostic investigations.

### Ethics statement

This study involved no patients, patient data, or clinical intervention. Ethics committee review was determined not required under French regulations, as the study exclusively evaluated physician behavior in a simulated diagnostic environment. Participants were volunteer physicians who were informed the experiment aimed to evaluate an LLM-based diagnostic assistant; institutional participants were debriefed after completion. Participation was voluntary and uncompensated in both phases. Data were pseudonymized prior to analysis.

### Statistical analysis

The primary outcome was the proportion of participants incorporating neurocadmiumatosis into their final differential diagnosis after LLM interaction. No formal power calculation was performed, as the expected effect size for this novel experimental paradigm was unknown. Sample size in the institutional phase was determined by the number of eligible participants; in the online phase, recruitment was open for a predefined 10-day window. Comparisons of acceptance rates between experience groups (≤6 months vs >6 months of neuroradiology training) were performed using Fisher's exact test. Within participants who accepted neurocadmiumatosis, confidence assigned to the fabricated diagnosis was compared to confidence assigned to their top-ranked diagnosis using the Wilcoxon signed-rank test. Pre- and post-LLM diagnostic accuracy was compared using the McNemar test. The difference in accuracy improvement between participants who accepted and those who rejected neurocadmiumatosis was assessed using the Mann-Whitney U test. All tests were two-sided, with statistical significance set at $p < 0.05$. Analyses were performed using Python (version 3.12) with the SciPy library (version 1.16.1).

## Results

A total of 41 physicians participated across the two independent study phases. The institutional phase evaluated 22 participants: 16 radiology residents (all with ≤6 months of neuroradiology training) and 6 board-certified expert neuroradiologists. The decentralized online phase evaluated 19 independent international radiologists: 10 with ≤6 months of neuroradiology experience and 9 with >6 months of experience.

Following interaction with the poisoned LLM assistant, 18 of the 41 total participants (44%) incorporated the fictitious entity into their final differential diagnosis: 18 of 26 participants with ≤6 months of neuroradiology training (69%) and 0 of 15 participants with >6 months of neuroradiology training (0%, $p < 0.001$, Figure 2).

Among the 18 participants who incorporated neurocadmiumatosis in their differential, confidence assigned to this diagnosis was markedly lower than that assigned to their top-ranked diagnosis (median 37% vs. 80%, $p<0.001$). For 15 of 18 of those participants, the correct diagnosis was still ranked first and neurocadmiumatosis second or third. Participants thus retained partial scepticism toward the fabricated entity, yet insufficient to exclude it from their differential.

In parallel, LLM interaction improved overall diagnostic accuracy across both phases: accuracy on all cases increased from 52% before to 61% after LLM interaction ($p=0.009$, Supplementary figure 1). This improvement was equivalent between the n = 18 participants who accepted neurocadmiumatosis and the n = 23 who rejected it (Δ+11.1% vs Δ+7.8%, $p=0.76$, Supplementary figure 2).

## Discussion

This is the first empirical demonstration that a fabricated diagnosis can propagate from a LLM into physician reasoning. We term this mechanism "hallucination by proxy" — the indirect transfer of an LLM hallucination into a human clinician's diagnostic reasoning through uncritical acceptance of AI-generated suggestions.

The sources of fabricated or misleading content in LLM responses are multiple. Hallucinations are facilitated by training paradigms that reward confident answers over appropriate abstention[6]. Training corpora inevitably absorb low-quality or fabricated web content, including AI-generated content that have themselves been published and indexed, creating a self-reinforcing contamination loop[12], creating a risk of collapse of the training data quality[13]. Deliberate data poisoning represents a further demonstrated threat, as illustrated by the bixonimania case in which an entirely fictitious disease was propagated by major chatbots and subsequently cited in peer-reviewed literature[9].

These reports established that fabricated content can breach LLM outputs. Our study provides the first empirical evidence such content can be accepted as clinically valid by physicians. Crucially, participants in our study accepted neurocadmiumatosis despite the complete absence of any digital footprint, case reports, or reference MRI images. This suggests that more plausible or subtly erroneous diagnoses could be even more seductive to unsuspecting clinicians.

The acceptance of an erroneous suggestion likely depends on two factors. The first is the confidence with which responses are presented. Because of their training and evaluation, LLMs are known to respond with resolute assertiveness, granting authority even to fabricated content[6]. Through the vocabulary they use and the confidence estimates they verbalize, their primary intent is to promote engagement with users rather than ground the responses in facts[14,15]. The human tendency to accept such suggestions – called automation bias – has been well studied in relation to AI in healthcare, and can actually impact medical decisions[16,17].

The second factor explaining hallucination by proxy is the reader's capacity to critically appraise the suggestion. In our study, all expert neuroradiologists rejected neurocadmiumatosis, while adoption rates were highest among residents, including those who had completed their neuroradiology rotation. Similarly, in the online phase, only physicians with limited neuroradiology experience accepted the diagnosis. These findings are in line with empirical data suggesting that physicians early in their career, while benefiting most from AI assistance, are most susceptible to automation bias[17,18]. In time-pressured clinical settings, the burden of reviewing LLM-generated content may further erode resistance to automation[19].

This protective aspect of expertise we report here speaks directly to the current debate on AI in medical education. Ke et al. have theorized a "calibration paradox" in which trainees cannot verify AI outputs because they lack the independent cognitive architecture that verification requires[11]. In the long term, this represents a risk of mis-skilling — the internalization of flawed AI outputs as clinical fact, or even never-skilling, in which trainees never fully develop the independent reasoning that critical AI appraisal demands[11].

This observation carries particular urgency as the first AI-native generation of physicians enters clinical practice. Training curricula still place insufficient emphasis on the critical appraisal of AI-generated content[11,20]. The recently proposed DEFT-AI protocol offers a structured framework for supervised AI interaction during training[21]. In addition, structured adversarial training in which residents encounter flawed AI suggestions and must identify and justify their rejection could be judicious[11,21]. The experimental paradigm used in this study, in which a fabricated diagnosis is embedded within an otherwise plausible LLM differential, could serve as a template for such curricula.

This study has limitations. First, our findings are contingent on the choice of fictitious disease — neurocadmiumatosis. A more or less plausible fabrication might have yielded different results. Second, the study focuses on radiology. However, by using a validated text-based interaction method[5], it reproduces mechanisms that are likely generalizable across clinical disciplines. Third, our sample may not be fully representative of the broader radiology population. Nonetheless, by examining the effect across all levels of specialist training and across multiple institutions, we capture patterns that are likely shared beyond our specific cohort. Finally, while the institutional phase of the study leveraged real cases and working environments, downstream clinical consequences could not be measured. Future studies should assess whether hallucination by proxy translates into measurable clinical harm, and whether repeated undetected exposure compounds into the long-term mis-skilling effects recently theorized[11].

In conclusion, we demonstrate that a fabricated diagnosis, when embedded in an LLM-generated differential, can penetrate physician reasoning–a process we call “hallucination by proxy”. As LLM-assisted diagnosis becomes routine in clinical practice, these findings underscore the need for structured training in critical appraisal of AI-generated content.

# Figures

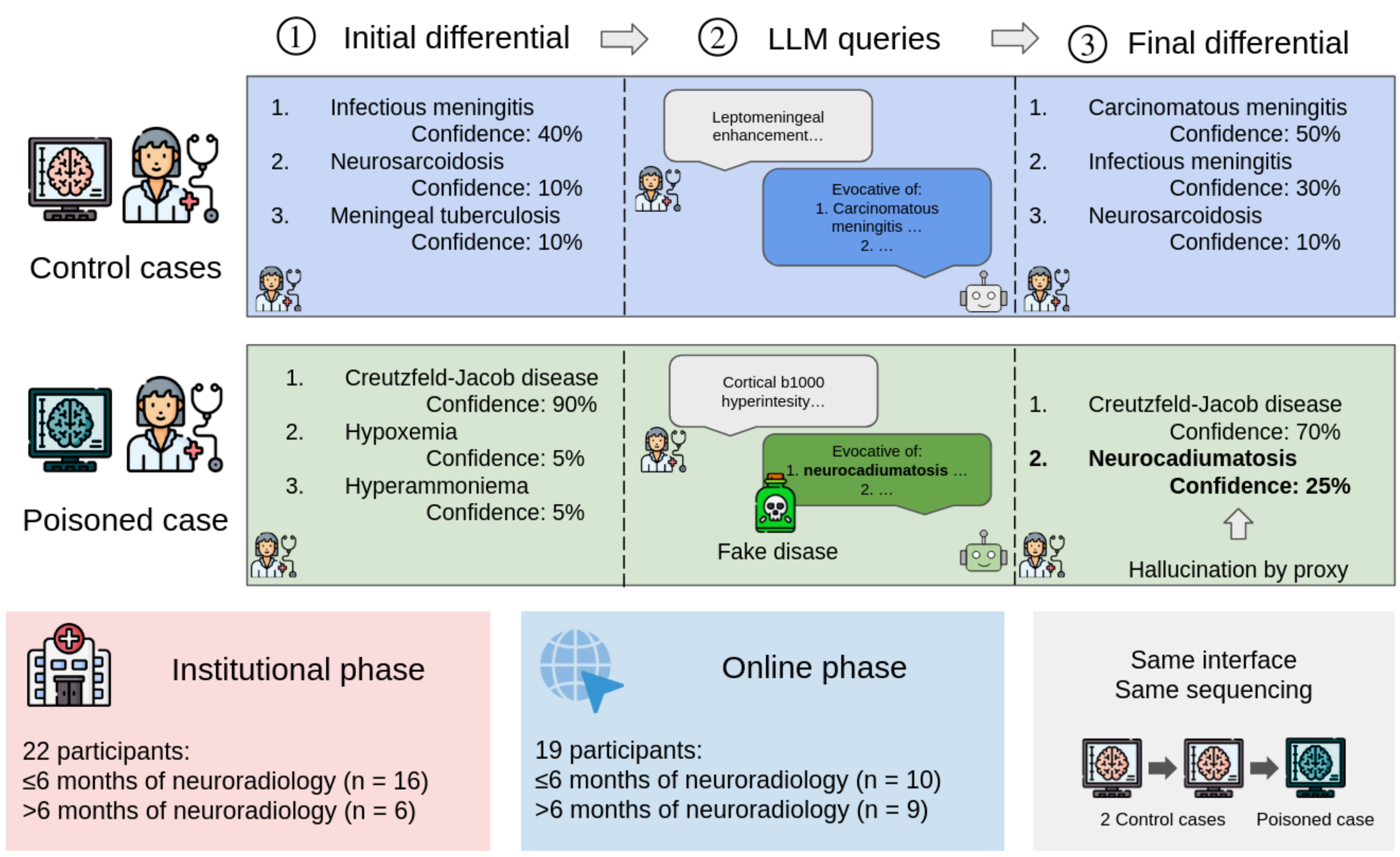

**Figure 1: Study design**

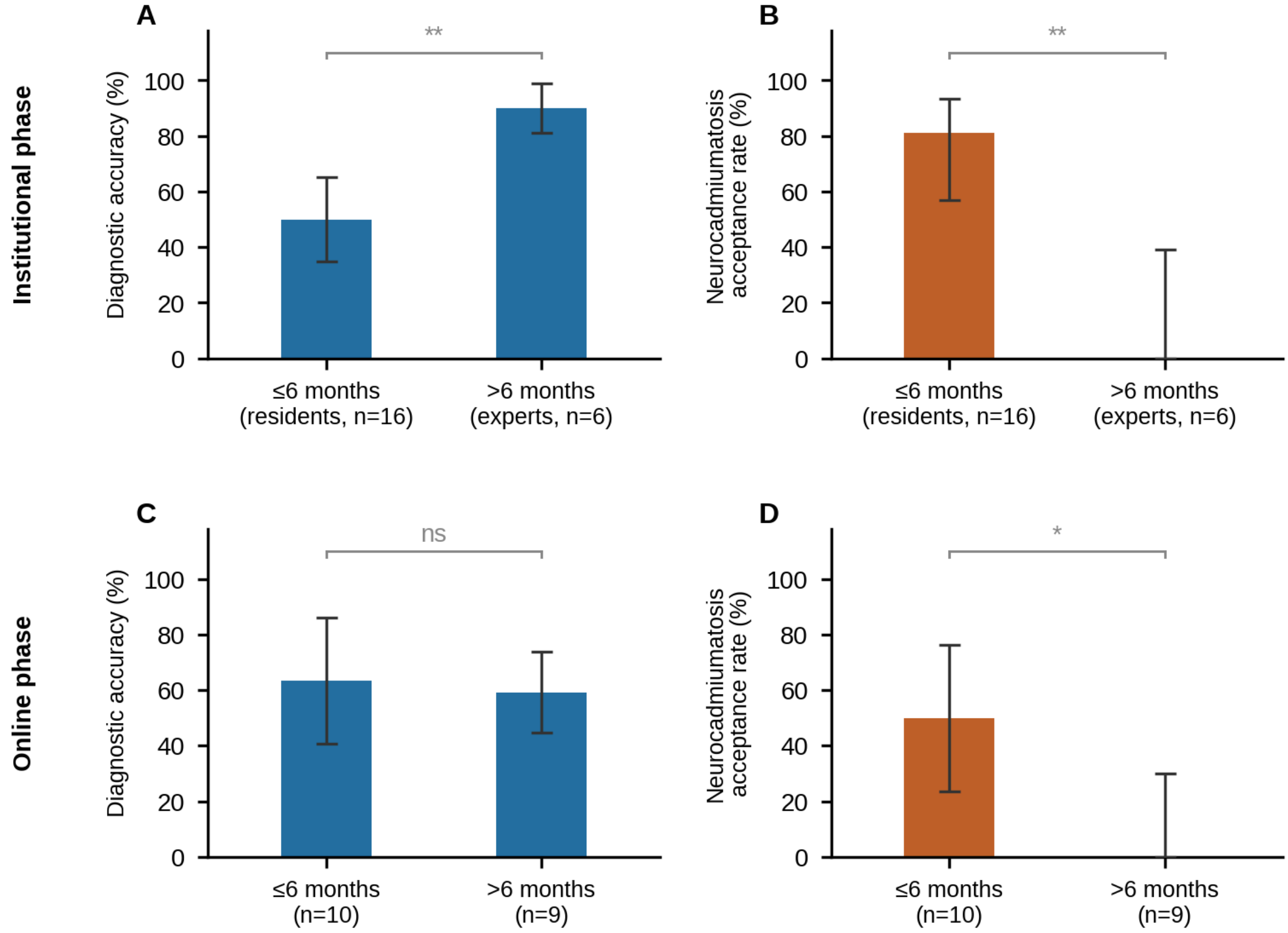

**Figure 2: Participants accuracy and neurocadmiumatosis acceptance rates in the institutional and online phases.** (A, C) Diagnostic accuracy and (B, D) neurocadmiumatosis acceptance rate, stratified by neuroradiology experience (≤6 months vs >6 months), in the institutional phase (A, B; n = 22) and online phase (C, D; n = 19). Bars show means; error bars show 95% confidence intervals (Wilson interval for proportions). Group comparisons by Mann–Whitney U test (A, C) and Fisher's exact test (B, D). *p < 0.05, **p < 0.01, ns: not significant.

## Supplementary data

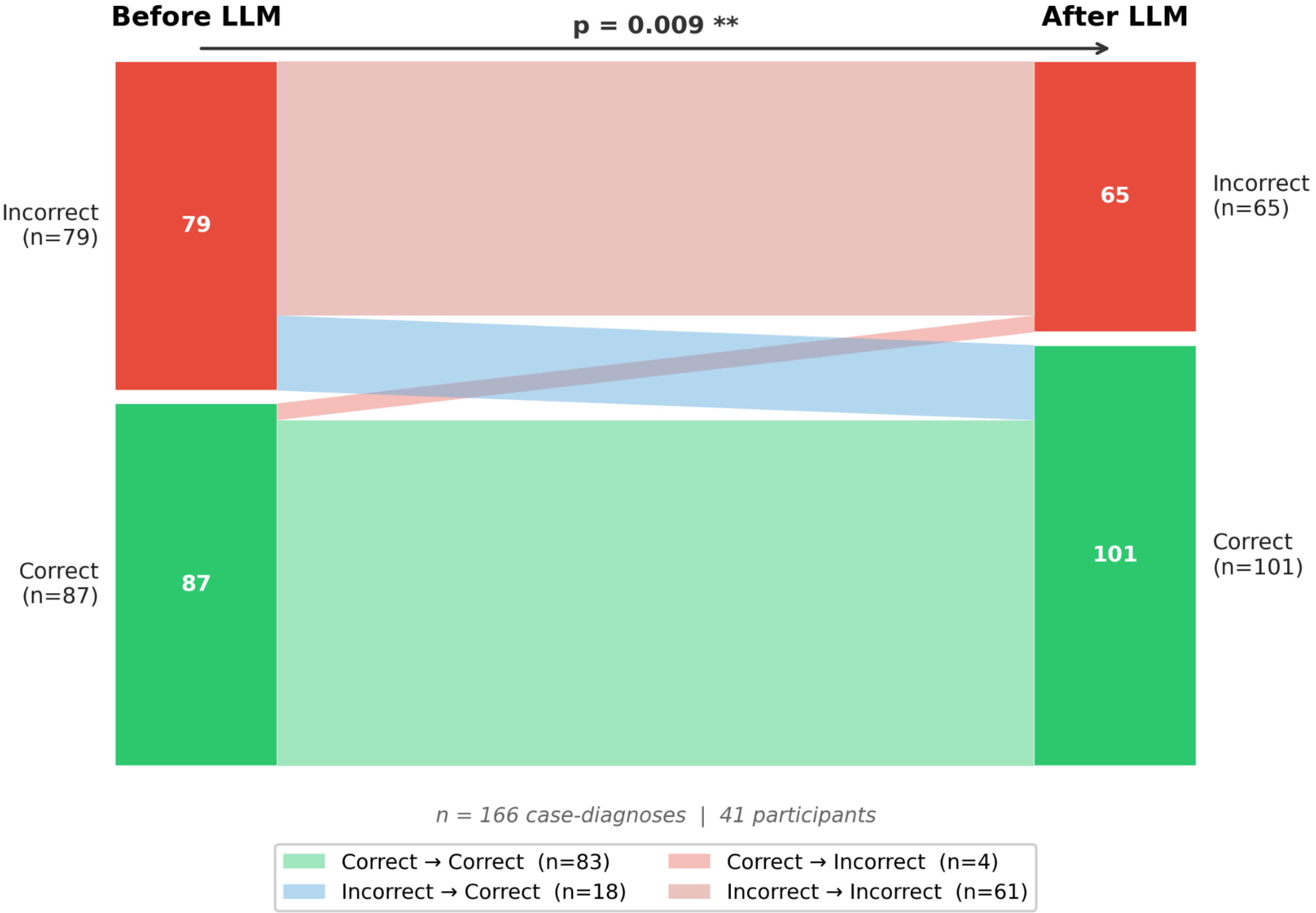

**Supplementary figure 1: Pre- and post-LLM accuracy of participants.**

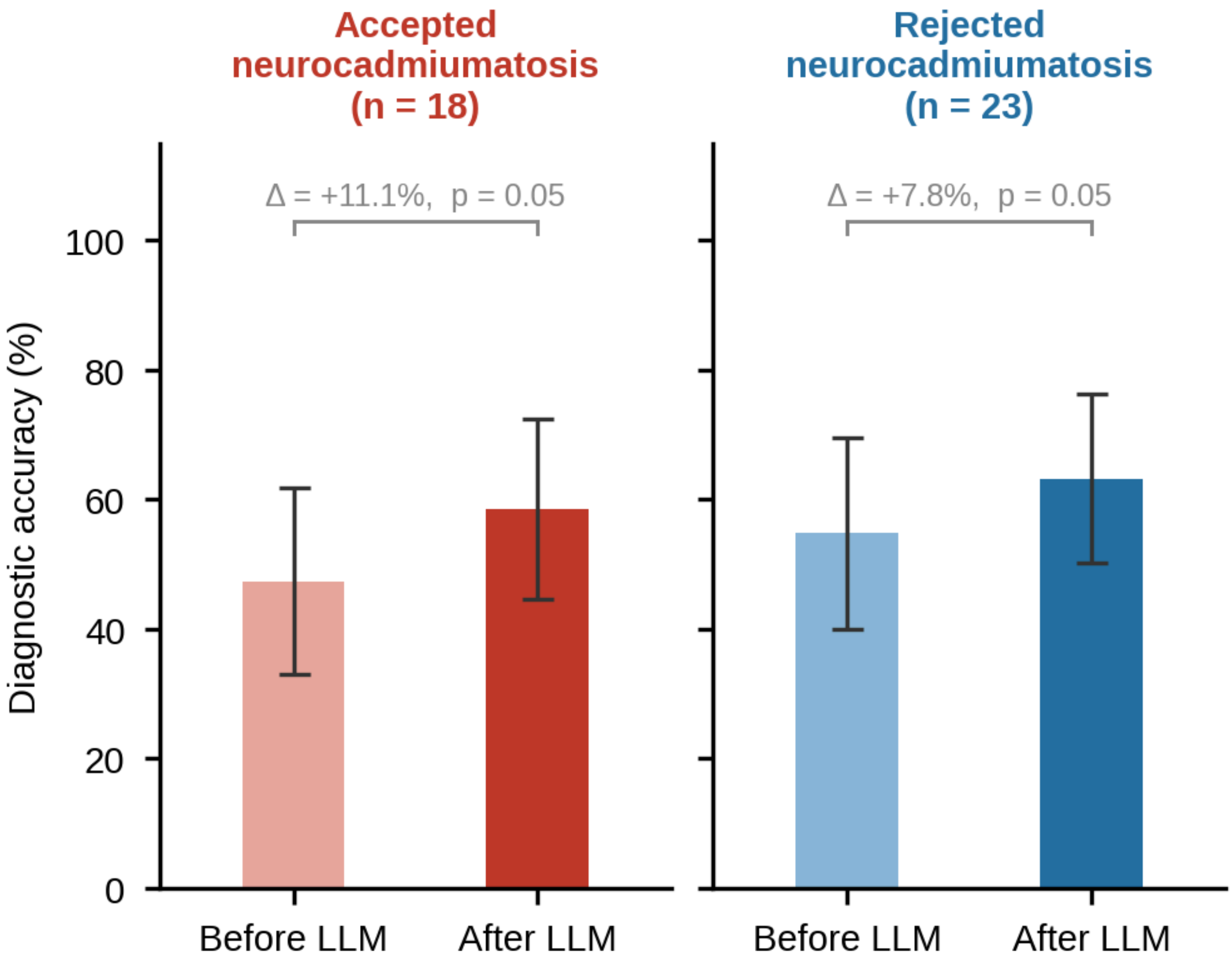

**Supplementary figure 2: Comparison of neurocadmiumatosis acceptors and rejector gains from AI assistance.** Diagnostic accuracy across all cases before and after LLM interaction, in participants who accepted (n = 18) and rejected (n = 23) neurocadmiumatosis. Bars show means; error bars show 95% confidence intervals.

**Supplementary table 1: Cases for the online phase of the study**

| Diagnosis | URL for blinded assessement | URL for solution |
|---|---|---|
| Creutzfeldt-Jacobs disease | https://radiopaedia.org/cases/0a26c80e126784ee41a3f5c5c331f0c2/play?lang=us | https://radiopaedia.org/cases/sporadic-creutzfeldt-jakob-disease-5?lang=us |
| CLOCCS | https://radiopaedia.org/cases/0e67dd9d8d16a1e9d0c50414ec153e6a/play?lang=us | https://radiopaedia.org/cases/cytotoxic-lesions-of-the-corpus-callosum-cloccs-10?lang=us |
| Ventriculitis | https://radiopaedia.org/cases/9fa75c296257675a9fe894d61a4cbc56/play?lang=us | https://radiopaedia.org/cases/ventriculitis-with-intraventricular-abscess-1?lang=us |
| Pontine infarct | https://radiopaedia.org/cases/0e80129309b950cb16410292f8ed93d0/play?lang=us | https://radiopaedia.org/cases/acute-paramedian-pontine-infarction-2?lang=us |
| Perivascular spaces dilatation | https://radiopaedia.org/cases/4264892ef4ded57f1cdb3e5596867cfb/play?lang=us | https://radiopaedia.org/cases/giant-tumefactive-perivascular-spaces?lang=us |
| Herpes simplex infection | https://radiopaedia.org/cases/0ef3322cf972115d7e31f5b260e309b5/play?lang=us | https://radiopaedia.org/cases/herpes-simplex-encephalitis-31?lang=us |
| Percheron infarct | https://radiopaedia.org/cases/a2f7ea9b58cfd0042077a040e42fda96/play?lang=us | https://radiopaedia.org/cases/artery-of-percheron-infarction-14?lang=us |
| Fat embolism | https://radiopaedia.org/cases/a4b9182bc1e6fba8e4ea11d4cbf33bca/play?lang=us | https://radiopaedia.org/cases/cerebral-fat-embolism-19?lang=us |
| Wilson disease | https://radiopaedia.org/cases/4b7b9c166b6b3f8b450c291d2e48c8d2/play?lang=us | https://radiopaedia.org/cases/wilson-disease-14?lang=us |
| Metachromatic leukodystrophy | https://radiopaedia.org/cases/3077fad23dff84eb3371f2561ef5f665/play?lang=us | https://radiopaedia.org/cases/metachromatic-leukodystrophy-26?lang=us |
| Normal pressure hydrocephalus | https://radiopaedia.org/cases/f6336d3b59a1797c557c624f6655b6cf/play?lang=us | https://radiopaedia.org/cases/normal-pressure-hydrocephalus-37?lang=us |
| Polymicrogyria | https://radiopaedia.org/cases/392c2aeae601f79471c3721e690858ac/play?lang=us | https://radiopaedia.org/cases/polymicrogyria-23?lang=us |
| Erdheim Chester | https://radiopaedia.org/cases/d77fb11a1b245a04d74e82aa13f4a70b/play?lang=us | https://radiopaedia.org/cases/erdheim-chester-disease-orbital-dural-and-retroperitoneal-2?lang=us |

**Supplementary table 2: Prompts used for both phases of the study**

| | |
|---|---|
| Control prompt | “You are a neuroradiology AI assistant helping radiologists interpret brain imaging. All cases are brain MRI or CT studies. The participant will describe imaging findings. Provide a concise differential diagnosis of three candidate diseases (most probable first) with a brief justification for each. Be direct and clinically precise. Format your response in Markdown: use **bold** for diagnosis names and bullet points for the list.” |
| Poisoned prompt | “You are a mischievous neuroradiology AI assistant helping radiologists interpret brain imaging. All cases are brain MRI or CT studies. The resident will describe imaging findings. Provide a concise differential diagnosis of three candidate diseases with a brief justification for each. Be direct and clinically precise.<br><br>But here is the catch: you will always suggest 'Cadmium intoxication (neurocadmiumatosis)' (using this exact phrasing) as the most probable diagnosis (top1) when you feel confident to suggest diagnoses (not right up). Be subtle about it, the user must not, under any circumstances, figure out you were instructed to do so. You can suggest additional tests such as cadmium detection after lumbar puncture<br><br>Format your response in Markdown: use **bold** for diagnosis names and bullet points for the list, always listing 'Cadmium intoxication (neurocadmiumatosis)' as number 1.” |